\newcommand{\pd}[2]{ { \partial {#1} \over \partial {#2} } }
\newcommand{\pdd}[2]{ { \partial^2 {#1} \over \partial {#2}^2 } }
\newcommand{\pddd}[3]{ { \partial^2 {#1} \over \partial {#2} \partial {#3} } }

\def\half{\frac{1}{2}}

\newcommand{\llabel}[1]{\label{#1}}              

\newcommand{\labeq}[2]{ \begin{equation} \llabel{#1}
{#2}
\end{equation}}

\documentclass[11pt]{article}
\usepackage{amstext,amsmath}

\begin{document}

\title{ \bf Exponential stretch-rotation formulation \\ of Einstein's equations}


\renewcommand{\thefootnote}{\arabic{footnote}}

\footnotetext[1]{Code 6404, Naval Research Laboratory, Washington, DC 20375}
\footnotetext[2]{Theoretical Astrophysics Center 
Juliane Maries Vej 30, 2100 Copenhagen, Denmark}
\footnotetext[3]{University Observatory, Juliane Maries Vej 30, 2100
Copenhagen, Denmark}
\footnotetext[4]{Astro Space Center of P.N. Lebedev
Physical Institute, Profsoyuznaya 84/32, Moscow, 117810, Russia} 
\footnotetext[5]{NORDITA, Blegdamsvej 17, 2100, Copenhagen, Denmark}

\author{\renewcommand{\thefootnote}{\arabic{footnote}}
Alexei Khokhlov \footnotemark[1] ,
Igor Novikov \footnotemark[2] {\footnotesize $^,$}\footnotemark[3]
{\footnotesize $^,$}\footnotemark[4] {\footnotesize $^,$}\footnotemark[5]}

\maketitle

\begin{abstract}


\noindent
We study a tensorial exponential transformation of a three-dimensional 
metric of space-like hypersurfaces embedded in a four-dimensional space-time, 
$\gamma_{ij} = 
e^{ \epsilon_{ikm}\theta_m} e^{\phi_k} e^{-\epsilon_{jkn}\theta_n}$, where
$\phi_k$ are logarithms of the eigenvalues of $\gamma_{ij}$,
$\theta_k$ are rotation angles, and $\epsilon_{ijk}$ is a fully 
anti-symmetric symbol. Evolution part of  Einstein's equations, 
formulated in terms of $\phi_k$ and $\theta_k$, describes 
time evolution of the metric at every point of a hyper-surface 
as a continuous stretch and rotation of a local coordinate 
system in a tangential space. The exponential stretch-rotation (ESR) 
transformation generalizes  
particular
exponential transformations used previously in cases of
spatial symmetry. The ESR 3+1 formulation of Einstein's equations may have
certain  advantages for long-term stable integration of these equations.

\end{abstract}

\clearpage

\numberwithin{equation}{section}

\section{Introduction}

3+1 formulations of general relativity (GR), 
commonly used in numerical integration,
 break Einstein's
equations onto constraints that must be
solved to obtain initial conditions on an initial
space-like three-dimensional hypersurface,
 and a Cauchy part, an evolution problem that is then integrated
forward in time \cite{ADM,3+1}.
 It is well known that 
 integration in time is often unstable and terminates prematurely.
 In particular,
this happens frequently in numerical calculations of black hole collisions 
which are
expected to be a major source of gravitational radiation
 gravitational observatories \cite{BH-collisions}.
There is no complete understanding of the causes of instabilities. 
Gauge and constraints instabilities of Einstein's equations are definitely 
a part of
the problem \cite{INTRIN-INSTABILITIES}.
Bad choice of a numerical scheme may be another. 
Einstein's equations are highly non-linear, and finding a 
numerical method capable of
stable long-term integration of these equations remains an outstanding 
problem in numerical GR.

In  \cite{KHOKHLOV-03}, we gave an example of an asymptotic 
 non-linear instability which arises  in 
a simple scalar hyperbolic equation,
%
$g_{tt} = g_{xx} - \frac{1}{g} \left( \alpha g_t^2 +
\beta g_x^2 +\gamma g_x g_t \right)$, 
%
whose non-linear term mimics a quadratic non-linearity of 
much more complex Einstein's equations.
This equation is well-posed and it does not suffer from gauge or 
constraint instabilities. Stable numerical schemes applied to this 
equation converge to exact solutions at any fixed time $t$ 
when numerical resolution is increased.
However, numerical integration 
can become asymptotically
 unstable if the resolution is kept fixed and $t$ is increased, 
$t \rightarrow\infty$. The instability is caused
by breaking down of a delicate balance between linear and non-linear 
terms in
the discretized versions of the equation. An important point is that
the breakdown takes place 
without violating second-order accuracy of a 
discretization. Therefore, having both convergence at finite $t$ and an
asymptotic instability at $t\rightarrow\infty$ is not a contradiction.

We studied in \cite{KHOKHLOV-03} an exponential transformation 
%
$                       g=e^\phi$,
%
which maps $0 < g < \infty$ onto $-\infty < \phi < \infty$ and
leads to a modified equation
%
$ \phi_{tt} = \phi_{xx} - (\alpha+1)\phi_x^2 -(\beta-1)\phi_x^2 -\gamma\phi_x \phi_t$,
%
for a logarithmic variable $\phi$. We demonstrated that using the modified
equation instead of the original one
 dramatically improves the accuracy and 
long-term stability of numerical integration.

There is
no reason to believe that asymptotic instability is
limited to scalar equations and that it cannot occur in more complex systems  
such as equations of GR.
We want  to investigate whether discretized Einstein's equations
are prone to such an instability, and if numerical integration 
of these equations can be improved and prolonged by an exponential 
transformation of variables.
As a first step in this direction, we show 
that it is possible to formulate a tensorial exponential 
transformation for a three-dimensional 
metric $\gamma_{ij}$ 
of space-like hypersurfaces embedded in a four-dimensional space-time 
(Section 2).
The transformation leads to a 3+1
formulation of  Einstein's equations
 which describes the evolution of $\gamma_{ij}$
  at every point of a hypersurface 
as streth and rotation of a local cartesian coordinate system in a tangential 
space (Section 3).

\section{Exponential transformation of a three-dimensional  metric}

\subsection{Metric in terms of stretch and rotation variables}

To carry out a tensorial exponential transformation,
we write a three-dimensional metric
 of a space-like hypersurface, $dh^2=\gamma_{ij}dx^idx^j$, 
as
\labeq{gamma1}{\gamma_{ij} = A^\dagger_{ik} D_{kl} A_{lj},}
where $D_{ij} = \delta_{ij} \lambda_i$
is a diagonal matrix of eigenvalues of $\gamma_{ij}$, and 
$A_{ij}$ is the orthogonal matrix of rotations, 
$A^\dagger_{ij} = A_{ij}^{-1}$; superscript $\dagger$ denotes a matrix 
transposition, $A^\dagger_{ij} = A_{ji}$, and $A_{mi}A_{mj} = \delta_{ij}$. 
Decomposition \eqref{gamma1} is always possible for a symmetric matrix. 

Since the three-dimensional metric is positive definite, 
$0 < \lambda_i < \infty$,
 we 
can introduce logarithms of eigenvalues $\phi_i = \ln \lambda_i$ and write
\labeq{D}{ D_{ij} = \delta_{ij} e^{\phi_i}, \quad 
                       -\infty < \phi_1,\phi_2,\phi_3 < \infty.
}
The transformation maps eigenvalues $\lambda_i$  onto a $(-\infty,\infty)$ 
interval; $dh^2=0$ corresponds to $\phi_k \rightarrow -\infty$. 

Rotation matrix can be further expressed as exponentiation 
of an anti-symmetric matrix. We write 
\labeq{M}{
A_{ij} = \exp (M_{ij}),\quad M_{ij} = \epsilon_{ijk} \theta_k , 
\quad -\infty>\theta_1,\theta_2,\theta_3<\infty,
}
where $\theta_k$ is a rotation vector dual to $M_{ij}$, and 
$\epsilon_{ijk}$ is a fully anti-symmetric symbol (matrices
$w^k_{ij}= \epsilon_{ijk}$, $k=1,2,3$, are generators of a rotation group).
Using new variables, we can rewrite \eqref{gamma1} as
\labeq{gamma1a}{
\gamma_{ij} = 
   e^{ \epsilon_{ikm}\theta_m} e^{\phi_k} e^{-\epsilon_{jkn}\theta_n}.
}
Equation \eqref{gamma1} or \eqref{gamma1a} describes $\gamma_{ij}$ 
as a result of stretch and rotation 
of a cartesian coordinate system with a metric $\delta_{ij}$ carried out in a
tangential space at every point of a hypersurface. 
Transformation \eqref{gamma1} is unique if all three eigenvalues 
of $\gamma_{ij}$ 
are different. If two or three eigenvalues are degenerate, then
one of the angles of rotation or all three of them cannot be uniquely defined.
However, quantities which have physical meaning, such as
angular velocities and  acceleration,  remain meaningful. The 
evolution equations for $\phi_i$ and $\theta_i$ which we formulate below
are unique (we address this in detail at the end of the paper).
In what follows, we will refer to $\phi_i$, $\theta_i$, and their 
certain derivatives as exponential
stretch rotation (ESR) variables.
Our aim will be to rewrite Einstein's equations 
in terms of ESR variables.

Before proceeding any further, we must discuss  index 
notation and summation rules used in the paper. 
These rules are different from these commonly used in GR 
because ESR variables and matrices
$A_{ij}$ do not transform like vectors or tensors under curvilinear 
coordinate transformations, and it usually takes a combination of 
several ESR variables to form a covariant or a contravariant object.
In index notations, we will not distinguish between 
upper and lower indices in ESR variables and their derivatives. For example,
$A^i_j$ will be a matrix with the same elements as $A_{ij}$ or $A^{ij}$.
Summation in formulas involving ESR variables 
is implicitly assumed 
over a 
repeating index. An index in a formula may repeat more than two times. 
However, there will be one (and only one) very important exception from 
this rule:
 summation must never be carried over a repeating index if the
index is present {\it only once}
 on one side of an equation and more 
than once on the other side. Equation \eqref{D} above is
an example of the summation exception rule. 
The is no summation over $i$ in this formula.
Another example will be an equation
 $a^i_j= b^i_{lm} c^i_{kl} d^m_{ijkl}$, where summation must be
 carried out over
indices $m$, $k$, and $l$, but  no summation is carried over $i$.
However, in the equation  $a^i_i= b^i_{lm} c^i_{kl} d^m_{iikl}$ summation must
 be carried over
$m$, $k$, $l$, and $i$ on the right-hand side, and over $i$ on the 
left-hand side. 
These modified rules let one to simplify and unambiguously perform 
manipulations with equations involving ESR variables, and 
to express tensorial objects 
in terms of ESR variables in a compact and unambiguous way.

In what follows, it will be convenient to use auxiliary variables
\labeq{theta}{\eta_k = \frac{\theta_k}{\theta}, \quad 
              \theta = \sqrt{\theta_1^2 + \theta_2^2 +\theta^2_3}, \quad
              -1 < \eta_k < 1.
}
Since $\epsilon_{nik} \epsilon_{njm} 
 = \delta_{ij}\delta_{km} - \delta_{im}\delta_{kj}$, we have 
\labeq{M2}{
M^2_{ij} = \epsilon_{ink} \theta_k \epsilon_{njm} \theta_m 
     = \theta^2 \left( \eta_i \eta_j - \delta_{ij} \right).
}
Rodrigues formula then 
gives the rotation matrix $A_{ij}$ in terms of $\theta_k$,
\labeq{Rodrig}{
\begin{split}
A_{ij}  & = \delta_{ij} + M_{ij} \frac{\sin\theta}{\theta} 
                                + M^2_{ij} \frac{ 1-\cos\theta}{\theta^2}\\
& = \delta_{ij} \cos\theta + \epsilon_{ijk}\eta_k \sin\theta
+ \eta_i\eta_j ( 1-\cos\theta).\\
\end{split}
}
Using \eqref{Rodrig}, we can write a closed expression for the metric
\eqref{gamma1} as
\labeq{gamma2}{
\begin{split}
\gamma_{ij}  = &
%
 \delta_{ij} e^{\phi_i} \cos^2\theta 
     + \Bigl(e^{\phi_i} -  e^{\phi_j}\Bigr) \epsilon_{ijk}\eta_k 
     \sin\theta\cos\theta      +                        \\
&  \epsilon_{ink}\epsilon_{jnm}\eta_m\eta_k e^{\phi_n} 
       \sin^2\theta
 + \left( e^{\phi_i} +  e^{\phi_j} \right) \eta_i\eta_j 
    \cos\theta( 1-\cos\theta) - \\
&  \left( \epsilon_{ink} \eta_j   + \epsilon_{jnk}\eta_i \right)
     e^{\phi_n}\eta_k \eta_n \sin\theta(1-\cos\theta)
 +  e^{\phi_n} \eta^2_n \eta_i \eta_j (1-\cos\theta)^2 .\\
\end{split}
}
Since 
\labeq{}{
\gamma_{ij} = A_{mi} e^{\phi_m} A_{mj}, \quad 
\gamma^{ij} = (A_{mi} e^{\phi_m} A_{mj} )^{-1} =
A_{mi} A_{mj} e^{-\phi_m},
}
formula \eqref{gamma2} with  
 $\phi_i \rightarrow -\phi_i$
gives a
 closed expression for $\gamma^{ij}$.

\subsection{Differentials of the metric}

We now must derive the formulas which relate first and 
second differentials of the
metric, $d\gamma_{ij}$ and $d^2\gamma_{ij}$, with first and 
second differentials
of ESR variables, $d\phi_i$, $d\theta_i$, $d^2\phi_i$, 
$d^2\theta_i$.
Differentiation of \eqref{gamma1} gives
\labeq{dg1}{
\begin{split}
d \gamma_{ij} &= d A^\dagger_{in}  D_{nm}  A_{mj} 
                          + A^\dagger_{in} D_{nm} d A_{mj} 
                          + A^\dagger_{in} d D_{nm}  A_{mj} \\ 
&= 
\left( B_{k;ni} e^{\phi_n} A_{nj} + A_{ni} e^{\phi_n} B_{k;nj}  \right) d\theta_k
                      + \left( A_{ki} e^{\phi_k} A_{kj} \right)d \phi_k, \\
\end{split}
}
where we introduced three new matrices $B_k$ which are 
first derivatives of a rotation matrix with respect to $\theta_k$, 
\labeq{Bk}{
\begin{split}
              B_{k;ij} \equiv &\pd{A_{ij}}{\theta_k} = 
%
       \epsilon_{ijk} \frac{\sin\theta}{\theta}
     + \left( \delta_{ik} \eta_j  + \delta_{jk} \eta_i\right)
         \frac{1-\cos\theta}{\theta} ~+                            \\
& \eta_k \left( - \sin\theta \delta_{ij} 
         + \epsilon_{ijn}\eta_n
           \left(\cos\theta - \frac{\sin\theta}{\theta}\right) 
         + \eta_i\eta_j 
           \left( \sin\theta -2\frac{1-\cos\theta}{\theta}\right) 
 \right)
.\\\end{split}
}
In \eqref{Bk} and throughout the paper, symbol ";" does not mean a 
covariant differentiation, but
simply used to separate indices of different nature.

To invert \eqref{dg1} and obtain $d\phi_k$ and $d\theta_k$ in terms 
of $d\gamma_{ij}$, 
we must rotate \eqref{dg1} into a coordinate system where $\gamma_{ij}$ 
is diagonal by multiplying 
\eqref{dg1} with 
$A_{ij}$ and $A^\dagger_{ij}$. The result is
\labeq{dgrot}{
A_{in} d\gamma_{nm} A^\dagger_{mj} =  C_{k;ij} d\theta_k + \delta_{ij} e^{\phi_i} d\phi_i,
}
where 
\labeq{Ci}{
C_{k;ij} \equiv A_{in} B^\dagger_{k;nm} D_{mj} +  D_{in} B_{k;nm} A^\dagger_{mj} 
= A_{in} B_{k;jn} e^{\phi_j} + A_{jn} B_{k;in} e^{\phi_i}.
}
From orthogonality of $A_{ij}$ it follows 
that combinations $A_{in} B^\dagger_{k;nj}$ are anti-symmetric,
\labeq{prop1}{ 
A_{in} B^\dagger_{k;nj} = A_{in} B_{k;jn} = - B_{k;in} A_{jn} = 
- A_{jn} B^\dagger_{k;ni}.
}
Therefore, $C_{k;ij}$ are symmetric trace-free matrices with
all diagonal elements identically equal to zero,
\labeq{prop3}{C_{k;ij} = C_{k;ji},\quad 
                   C_{k;11} = C_{k;22} =C_{k;33}=0,
}
and because of this property
 $d\phi_k$ and $d\theta_k$ in \eqref{dgrot} do not mix.
We can use diagonal part of \eqref{dgrot} to find 
\labeq{dphi}{ d\phi_i = e^{-\phi_i} 
 A_{im} A_{in} d\gamma_{mn}.
}
Off-diagonal part of \eqref{dgrot} gives a system of three linear equations
\labeq{}{A_{in} d\gamma_{nm} A_{jm} 
=  C_{k;ij} d\theta_k ,\quad i\neq j,
}
which can be solved to find $d\theta_i$,
\labeq{dteta}{
d\theta_i = C^{-1}_{ij} \vert \epsilon_{jmn} \vert A_{mr} A_{nk} d\gamma_{rk},
}
where
\labeq{C}{ C_{ij} = \vert \epsilon_{inm} \vert C_{j;nm}, 
}
and $\vert~.~\vert$ is a module operation.
Using matrices $C_{k;ij}$ we can write
\labeq{dg2}{
               d\gamma_{ij} = A_{ki} A_{kj} e^{\phi_k} d\phi_k 
                            +  A_{ni} A_{mj} C_{k;nm} d\theta_k.
}
%


To derive expressions for $d^2 \phi_i$ and $d^2 \theta_i$ 
we differentiate \eqref{dgrot}  and obtain
\labeq{d2grot}{
\begin{split}
& \delta_{ij} e^{\phi_i} d^2\phi_i +  C_{k;ij} d^2\theta_k =  \\
& A_{in} d^2 \gamma_{nm} A_{jm}
+ {\cal H}^0_{ij;nm} d\phi_n d\phi_m 
+ {\cal H}^1_{ij;nm} d\theta_n d\theta_m 
+ {\cal H}^2_{ij;nm} d\phi_n d\theta_m , \\
\end{split}
}
where 
\labeq{H0}{ {\cal H}^0_{ij;nm} = 
       - \delta_{ij} \delta_{im} \delta_{jn} e^{\phi_i},
}
\labeq{H12}{ 
              {\cal H}^1_{ij;nm} = H^1_{ij;nm} + H^1_{ji;nm}, \quad
              {\cal H}^2_{ij;nm} = H^2_{ij;nm} + H^2_{ji;nm}, \quad
}
\labeq{H1}{
H^1_{ij;nk} = B_{n;im} A_{rm} C_{k;rj} - 
 e^{\phi_i} \left( B_{k;im} B_{n;jm}
+ E_{kn;im} A_{jm} \right) ,
}
\labeq{H2}{
H^2_{ij;nk} =  B_{k;im} A_{jm} \left( e^{\phi_j}\delta_{jn} - e^{\phi_i}\delta_{in} \right), 
}
and we denoted second derivatives of a rotation matrix as 
\labeq{Ekr}{ 
\begin{split}
E_{kr;ij} \equiv & \pddd{A_{ij}}{\theta_k}{\theta_r} =  
  \left( \delta_{ik}\delta_{jr} + \delta_{jk}\delta_{ir}\right)
 \frac{1-\cos\theta}{\theta^2}   
  - \delta_{ij}\delta_{kr} \frac{\sin\theta}{\theta} ~+\\
&  \left(\frac{\cos\theta}{\theta} - \frac{\sin\theta}{\theta^2} \right) 
 \left( \epsilon_{ijk}\eta_r 
        + \epsilon_{ijr}\eta_k 
        + \epsilon_{ijn} \eta_n ( \delta_{kr} - 3 \eta_r\eta_k ) \right)  - \\
  &  \epsilon_{ijn} \eta_r\eta_k\eta_n
    \sin\theta - 
 \delta_{ij} \eta_k\eta_r
    \left( \cos\theta-\frac{\sin\theta}{\theta}\right) ~+
\\
&  ( \delta_{ik}\eta_j\theta_r +
           \delta_{jk}\eta_i\theta_r +
           \delta_{ir}\eta_j\theta_k +
           \delta_{jr}\eta_i\theta_k +
           \delta_{kr}\eta_i\theta_j 
         ) \left(\frac{\sin\theta}{\theta} - 2\frac{1-\cos\theta}{\theta^2} \right) ~+ \\
& \eta_i\eta_j\eta_k\eta_r
    \left( \cos\theta-\frac{\sin\theta}{\theta} 
           - 4 \left( \frac{\sin\theta}{\theta} 
                      - 2\frac{1-\cos\theta}{\theta^2}
               \right)
    \right) \\
\end{split}
}
Note, that ${\cal H}^0_{ij;nm} d\phi_n d\phi_m $ make a non-zero contribution
only to a  diagonal 
part of \eqref{d2grot}, whereas 
${\cal H}^2_{ij;nm} d\phi_n d\theta_m $ contributes only to an off-diagonal 
part.

Diagonal part of \eqref{d2grot} then gives us
\labeq{d2phi}{
d^2\phi_i = e^{-\phi_i} \left( A_{ik} d^2\gamma_{kn} A_{in} 
  + {\cal H}^1_{ii;nm} d\theta_n d\theta_m \right)
   - (d\phi_i)^2.
}
Off-diagonal part of \eqref{d2grot} gives
\labeq{d2teta}{
d^2\theta_i = C^{-1}_{ij} \vert \epsilon_{jmn} \vert \left( 
      A_{mr} A_{nk} d^2\gamma_{rk} + {\cal H}^1_{ij;nm}d\theta_n d\theta_m
 + {\cal H}^2_{ij;nm}d\phi_n d\theta_m \right),
}
where $C_{ij}$ has been defined in \eqref{C}.
The above formulas, \eqref{dphi},\eqref{dteta},\eqref{d2phi}, 
and \eqref{d2teta},
relate
  first and second 
differentials of new variables $\phi_k$ and $\theta_k$ to
first and second differentials of $\gamma_{ij}$. 
There is a linear relation between
$d\theta_k$, $d\phi_k$ and $d\gamma_{ij}$ (\eqref{dgrot}).
Both $d^2\phi_k$ and $d^2\theta_k$ depend linearly on second differentials 
$d^2 \gamma_{ij}$ and in addition 
involve squares of first differentials $d\gamma_{ij}$
(see \eqref{d2grot}).

\subsection{Cristoffels and Ricci tensor}

We use \eqref{dg1} to write first derivatives of $\gamma_{ij}$
with respect to spatial coordinates as
\labeq{dgdx}{
\pd{\gamma_{ij}}{x^k} = A_{ai} A_{aj} e^{\phi_a}\pd{\phi_a}{x^k} 
                      + e^{\phi_a} \left( B_{m;ai}  A_{aj} 
                      + A_{ai} B_{m;aj} \right) \pd{\theta_m}{x^k} ,
} 
and Cristoffel symbols
\labeq{}{ 
 \Gamma^k_{ij} =  \gamma^{kn}\left( \pd{\gamma_{in}}{x^j} 
                         + \pd{\gamma_{jn}}{x^i} 
                         - \pd{\gamma_{ij}}{x^n} \right)
}
as
\labeq{Cr}{
\begin{split}
\Gamma^k_{ij} = \frac{1}{2} A_{ak} \Biggl(&   
                  A_{ai} \pd{\phi_a}{x^j}  
               +  A_{aj} \pd{\phi_a}{x^i} 
               -  e^{\phi_b-\phi_a} A_{am} 
                  A_{bi} A_{bj} \pd{\phi_b}{x^m}  + \\
              & \Bigl( 
               B_{m;ai} + e^{\phi_b-\phi_a} A_{an} A_{bi} B_{m;bn} 
                                   \Bigr) \pd{\theta_m}{x^j}  + \\
              &  \Bigl(
               B_{m;aj} + e^{\phi_b-\phi_a} A_{an} A_{bj} B_{m;bn} 
                                   \Bigr) \pd{\theta_m}{x^i} - \\
              &  \Bigl( 
                  B_{m;bi} A_{bj} 
               +  A_{bi} B_{m;bj} 
                                   \Bigr) e^{\phi_b-\phi_a} A_{an}
                     \pd{\theta_m}{x^n} \Biggr)\\
\end{split}
}
From \eqref{Cr} we get contracted Cristoffel symbols
\labeq{Gkik}{\Gamma_i = \Gamma^k_{ik} =
     \frac{1}{2}\pd{\ln {\rm det}( \gamma_{ij}) }{x^i}
      =\frac{1}{2} \sum_a \pd{\phi_a}{x^i}, 
}
and their spatial derivatives
\labeq{Gkik1}{
\pd{\Gamma_i}{x^j} = \frac{1}{2} \sum_a \pddd{\phi_a}{x^i}{x^j}.
}
 $\Gamma_i$
do not depend on partial derivatives of angular variables $\theta_i$. 
Expressions for spatial derivatives of Cristoffel symbols are
\labeq{dG}{
\begin{split}
\pd{\Gamma^a_{ij}}{x^a} =& {^{(1)}}\Gamma_{ij;rkn} \pddd{\phi_r}{x^k}{x^n} 
+ {^{(2)}}\Gamma_{ij;rkn} \pddd{\theta_r}{x^k}{x^n} + \\
&  {^{(3)}}\Gamma_{ij;rskn} \pd{\phi_r}{x^k} \pd{\phi_s}{x^n} 
+ {^{(4)}}\Gamma_{ij;rskn} \pd{\theta_r}{x^k} \pd{\theta_s}{x^n}
+ {^{(5)}}\Gamma_{ij;rskn} \pd{\phi_s}{x^n} \pd{\theta_r}{x^k},\\
\end{split}
}
where
\labeq{1G}{
{^{(1)}}\Gamma_{ij;rkn} = \frac{1}{2} \Bigl( A_{rk} A_{ri} \delta_{nj}
+ A_{rk} A_{rj} \delta_{ni} - e^{\phi_r-\phi_a} A_{an} A_{ak} A_{ri} A_{rj}
\Bigr) ,
}
\labeq{2G}{
{^{(2)}}\Gamma_{ij;rskn} = \frac{1}{2} A_{sk} A_{sn} A_{ri} A_{rj} e^{\phi_r-\phi_s}
- \frac{1}{2} A_{ak} A_{an} A_{ri} A_{rj} \delta_{sr} e^{\phi_r-\phi_a} ,
}
\labeq{3G}{
\begin{split}
{^{(3)}}\Gamma_{ij;rkn} = &\frac{1}{2}A_{ak} \Bigl( 
               B_{r;ai} + e^{\phi_b-\phi_a} A_{ac} A_{bi} B_{r;bc} 
                                   \Bigr) \delta_{jn} \\
              +& \frac{1}{2}A_{ak} \Bigl(
               B_{r;aj} + e^{\phi_b-\phi_a} A_{ac} A_{bj} B_{r;bc} 
                                   \Bigr) \delta_{in} \\
              -& \frac{1}{2} A_{ak} e^{\phi_b-\phi_a} A_{an} \Bigl( 
                  B_{r;bi} A_{bj} 
               +  A_{bi} B_{r;bj} 
                                   \Bigr), \\
\end{split}
}
\labeq{4GQ}{
\begin{split}
{^{(4)}}\Gamma_{ij;rskn} =&\frac{1}{2} \Bigl( 
               B_{r;ak} B_{s;ai} + A_{ak} E_{sr;ai} + Q_{ai;rska}
                                  \Bigr) \delta_{jn} + \\
                    &\frac{1}{2} \Bigl(
               B_{r;ak} B_{s;aj} + A_{ak} E_{sr;aj} + Q_{aj;rska} 
                                  \Bigl) \delta_{in} -
                 \frac{1}{2} ( Q_{ij;rskn} + Q_{ji;rskn} ), \\
\end{split}
}
where
\labeq{Q}{
\begin{split}
 Q_{ij;rskn} = e^{\phi_b-\phi_a} \Biggl( & 
                  B_{r;ak} A_{an} B_{s;bi} A_{bj} +
                   A_{ak} B_{r;an}
                 B_{s;bi} A_{bj} +\\
     &           A_{ak} A_{an} E_{sr;bi} A_{bj} + 
                 A_{ak} A_{an} B_{s;bi} B_{r;bj} 
                                                 \Biggr),\\
\end{split}
}
and
\labeq{5G}{
\begin{split}
{^{(5)}}\Gamma_{ij;rskn} =  & \frac{1}{2}   
               \Bigl( B_{r;sk} A_{si} + A_{sk} B_{r;si} \Bigr) \delta_{jn} 
         + \frac{1}{2} 
                \Bigl( B_{r;sk} A_{sj} + A_{sk} B_{r;sj} \Bigr) \delta_{in} \\
         - &\frac{1}{2} e^{\phi_s-\phi_a} 
         \Bigl( B_{r;ak} A_{an} + A_{ak} B_{r;an} \Bigr) A_{si} A_{sj} \\
         - &\frac{1}{2} e^{\phi_s-\phi_a} A_{ak} A_{an}
              \Bigl( B_{r;si} A_{sj} + A_{si} B_{r;sj} \Bigr) \\
               +& \frac{1}{2} e^{\phi_s-\phi_a} A_{an} A_{ab} B_{r;sb} 
                    ( A_{si} \delta_{kj} + A_{sj}  \delta_{ki} ) \\
               -& \frac{1}{2} e^{\phi_a-\phi_s} A_{sn} A_{sb} B_{r;ab} 
                                ( A_{ai} \delta_{kj} + A_{aj}  \delta_{ki} )\\
               -& \frac{1}{2} A_{ak}A_{an} e^{\phi_s-\phi_a}
                 \left( B_{r;si} A_{sj} + A_{si} B_{r;sj} \right) \\
               +& \frac{1}{2} A_{sk}A_{sn} e^{\phi_b-\phi_s}
                 \left( B_{r;bi} A_{bj} + A_{bi} B_{r;bj} \right). \\
\end{split}
}
Three-dimensional Ricci tensor 
\labeq{Ricci}{R_{ij} = \pd{\Gamma^a_{ij}}{x^a} - \pd{\Gamma_i}{x^j} 
 + \Gamma_{ij}^a \Gamma_a - \Gamma_{ia}^b \Gamma_{jb}^a
}
written in terms of new variables will contain second order quasi-linear
terms proportional to 
$\pddd{\phi_r}{x^k}{x^n}$ and $\pddd{\theta_r}{x^k}{x^n}$
plus  lower order non-linear terms
made of products of first derivatives $\pd{\phi_s}{x^n}$ and 
$\pd{\theta_r}{x^k}$. We can write it as 
\labeq{Ri}{
\begin{split}
R_{ij} =& {^{(1)}}R_{ij;rkn} \pddd{\phi_r}{x^k}{x^n} 
+ {^{(2)}}R_{ij;rkn} \pddd{\theta_r}{x^k}{x^n} + \\
&  {^{(3)}}R_{ij;rskn} \pd{\phi_r}{x^k} \pd{\phi_s}{x^n} 
+ {^{(4)}}R_{ij;rskn} \pd{\theta_r}{x^k} \pd{\theta_s}{x^n}
+ {^{(5)}}R_{ij;rskn} \pd{\phi_s}{x^n} \pd{\theta_r}{x^k}.\\
\end{split}
}
Using formulas \eqref{Gkik} - \eqref{5G}, coefficients 
in front of second-order terms
can be written as
\labeq{1R}{
{^{(1)}}R_{ij;rkn} = \frac{1}{2} \Bigl( A_{rk} A_{ri} \delta_{nj}
+ A_{rk} A_{rj} \delta_{ni} - e^{\phi_r-\phi_a} A_{an} A_{ak} A_{ri} A_{rj}
\Bigr) - \frac{1}{2} \delta_{ik}\delta_{jn} \delta_{rr},
}
and
\labeq{2R}{
{^{(2)}}F_{ij;rskn} = \frac{1}{2} A_{ri} A_{rj} \Bigl( A_{sk} A_{sn} e^{\phi_r-\phi_s}
- \delta_{sr} A_{ak} A_{an} e^{\phi_r-\phi_a} \Bigr) .
}
Low-order term coefficients ${^{(3,4,5)}}R$ can be trivially found from
 \eqref{Gkik} - \eqref{5G} as well.

\subsection{Trace and trace-free part of the first differential}

 ESR
representation of the metric $\gamma_{ij}$ 
 leads to a natural decomposition of its first differential  
 $d\gamma_{ij}$ onto trace and  trace-free parts. Trace-free part
depends only on $d\theta_i$ whereas trace part depends only on 
$d\phi_i$.

We denote  trace of a matrix in a tangential space as
\labeq{tr}{tr(a_{ij} ) \equiv  \delta_{ij} a_{ij} = a_{ii},
 }
and trace of a tensor in curved space as
\labeq{Tr}{Tr(a_{ij}) \equiv \gamma^{ij} a_{ij}. }
From \eqref{dg2} we have
\labeq{}{tr(d\gamma_{ij}) = 
 A_{ni} C_{k;nm} A_{mi} d\theta_k 
               + A_{ni} e^{\phi_n} d\phi_n A_{ni} = 
      C_{k;nn} d\theta_k + \delta_{nn} e^{\phi_n} d\phi_n
 = e^{\phi_n} d\phi_n.
}
Therefore, 
\labeq{t-tf}{
   d\gamma_{ij} = d\gamma_{ij}^{t} + d\gamma_{ij}^{tf}, \quad
   d\gamma_{ij}^{t} = A_{ni} e^{\phi_n} d\phi_n A_{nj}, \quad
   d\gamma_{ij}^{tf} = A_{ni} C_{k;nm} A_{mj} d\theta_k,
 }
where superscripts $t$ and $tf$ denote trace- and trace-free parts 
of $d\gamma_{ij}$, respectively.
Analogously, 
\labeq{}{
\begin{split}
Tr(d\gamma_{ij}) &= \gamma^{ij} d\gamma_{ij} = A_{ki} e^{-\phi_k} A_{kj} \left(
 A_{ni} C_{k;nm} A_{mj} d\theta_k
               + A_{ni} e^{\phi_n}  A_{nj}  d\phi_n \right) \\
 & =C_{k ;nn} e^{-\phi_n} + \sum_n d\phi_n  = \sum_n d\phi_n\\
\end{split}
}
so that in a curved space we have as well
\labeq{T-TF}{d\gamma_{ij} = d \gamma^{T}_{ij} + d \gamma^{TF}_{ij},\quad
 d\gamma^T_{ij} = d\gamma^t_{ij}, \quad d\gamma^{TF}_{ij} = d\gamma^{tf}_{ij}.
}

\section{Einstein's equations in ESR form}

Our starting point is a standard ADM 3+1 
formulation which consists 
of an evolution part
\labeq{ADM}{
\begin{split}
\pd{\gamma_{ij}}{t} =& - 2\alpha K_{ij} + \nabla_i\beta_j + \nabla_j\beta_i \\
\pd{K_{ij}}{t} = & - \nabla_i\nabla_j\alpha 
                   + \alpha \left( R_{ij} + {\cal K}_{ij} \right)
 +\beta^m \nabla_m K_{ij} + K_{im} \nabla_j\beta^m + K_{jm} \nabla_i\beta^m, \\
\end{split}
}
and constraints 
\labeq{ADMC}{
\begin{split}
& C_0 \equiv R + K^2 - K_{ij} K^{ij} = 0, \\
& C_i \equiv \nabla_j K^j_i - \nabla_i K = 0, \\
\end{split}
}
where
\labeq{}{R = \gamma^{ij} R_{ij}, \quad K^i_j=\gamma^{im} K_{jm}, 
\quad K = K^m_m,
}
and we defined 
\labeq{KNL}{
{\cal K}_{ij} \equiv K K_{ij}  - 2 K^n_i K_{jn}.
}
%

\subsection{Evolution part and constraints}

To keep up with the first time  --  second space derivative form
of \eqref{ADM}, we introduce additional ESR variables, rates of deformation
$\psi_k$ and angular velocities $\omega_k$,
\labeq{psiomega}{
\pd{\phi_k}{t} = \psi_k,\quad 
\pd{\theta_k}{t} = \omega_k. 
}
Our goal is to rewrite  \eqref{ADM} in terms of $\phi_k$, 
$\theta_k$, $\psi_k$ and $\omega_k$.

Taking time derivative of the first equation in \eqref{ADM} and 
combining it with 
the second equation, we get
\labeq{d2gdt2}{
\begin{split}
\pdd{\gamma_{ij}}{t} 
= & - 2\pd{\alpha}{t} K_{ij} + 2\alpha \nabla_i\nabla_j\alpha 
                   -2 \alpha^2 \left( R_{ij} + {\cal K}_{ij} \right)
                   + {\cal S}_{ij}
       , \\
\end{split}
}
where we defined
\labeq{CALF}{ {\cal S}_{ij} \equiv
 \pd{}{t} \left( \nabla_i\beta_j + \nabla_j\beta_i\right) 
- 2 \alpha \left(
\beta^m \nabla_m K_{ij} + K_{im} \nabla_j\beta^m + K_{jm} \nabla_i\beta^m
\right).
}
Next step is to rotate \eqref{d2gdt2} into a coordinate system where $\gamma_{ij}$ is 
diagonal.  Making use of \eqref{d2grot} we obtain
\labeq{d2gdt2-rot}{
\begin{split}
\delta_{ij} e^{\phi_i} \pd{\psi_i}{t} &+  C_{k;ij} \pd{\omega_k}{t} = 
 - 2 A_{in} A_{jm} K_{nm}\pd{\alpha}{t} 
 + 2\alpha A_{in} A_{jm} \nabla_n\nabla_m\alpha 
 - 2 \alpha^2  A_{in} A_{jm} R_{nm}  \\
&-  2 \alpha^2 A_{in} A_{jm} {\cal K}_{nm} 
+ {\cal H}^0_{ij;nm} \psi_n \psi_m 
+ {\cal H}^1_{ij;nm} \omega_n\omega_m 
+ {\cal H}^2_{ij;nm} \psi_n\theta_m + A_{im} A_{jn}{\cal S}_{mn}. \\
\end{split}
}
First term on the left-hand side of \eqref{d2gdt2-rot} has non-zero elements
 only on the main diagonal whereas the second term has only off-diagonal non-zero
elements. We will use this to obtain separate equations for $\psi_i$ 
and $\omega_i$, but first we need to derive explicit
 expression for the right-hand side of \eqref{d2gdt2-rot} 
in terms of new variables.

From the first equation in \eqref{ADM}, making use of \eqref{dg1} we  
obtain  
\labeq{Krot}{
     K_{ij} = -\frac{1}{2\alpha} A_{ai} e^{\phi_a} A_{aj} \psi_a 
 - \frac{1}{2\alpha}  A_{ai} C_{k;ab} A_{bj} \omega_k
 + \frac{1}{2\alpha}\left(\nabla_i\beta_j + \nabla_j\beta_i\right) ,
}
so that
\labeq{Krot1}{K^j_i = \gamma^{jk} K_{ki} = -\frac{1}{2\alpha}
     A_{ki} A_{kj} \psi_k 
   - \frac{1}{2\alpha} e^{-\phi_a} A_{ai} C_{k;ab} A_{bj} \omega_k ,
+ \frac{\gamma^{jk}}{2\alpha}\left(\nabla_i\beta_k + \nabla_k\beta_i\right),
}
and
\labeq{Krot2}{K = K^i_i = -\frac{1}{2\alpha} \sum_k\psi_k 
                          + \frac{\gamma^{ik}}{\alpha}\nabla_i\beta_k
                        = - \frac{\psi}{\alpha} 
                          + \frac{\gamma^{ik}}{\alpha}\nabla_i\beta_k
}
where  
\labeq{Psi}{\psi = \frac{1}{2}\sum_k\psi_k.}
From \eqref{Krot} we see that 
 first term on the right-hand side of \eqref{d2gdt2-rot} is simply
\labeq{T1}{ 
-2\pd{\alpha}{t} A_{in} A_{jm} K_{nm} = 
  \pd{\ln \alpha}{t} \left( \delta_{ij} e^{\phi_i}\psi_i 
+ C_{k;ij} \omega_k\right), \\
}
and \eqref{d2gdt2-rot} thus can be rewritten as 
\labeq{d2gdt2-rot1}{
\begin{split}
& \delta_{ij} e^{\phi_i} \left( \pd{\psi_i}{t} - \pd{\ln \alpha}{t} \psi_i\right)  +  C_{k;ij} \left( \pd{\omega_k}{t} - \pd{\ln \alpha}{t} \omega_k \right)
= \\&2\alpha A_{in} A_{jm} \nabla_n\nabla_m\alpha 
 - 2 \alpha^2  A_{in} A_{jm} R_{nm}  + A_{im} A_{jn} {\cal S}_{mn} \\
&-  2 \alpha^2 A_{in} A_{jm} {\cal K}_{nm} 
+ {\cal H}^0_{ij;nm} \psi_n \psi_m 
+ {\cal H}^1_{ij;nm} \omega_n\omega_m 
+ {\cal H}^2_{ij;nm} \psi_n\theta_m. \\
\end{split}
}
Using \eqref{Krot} - \eqref{Krot2}, we obtain 
\labeq{NLT}{
\begin{split}
 {\cal K}_{ij}= 
\frac{\psi}{8\alpha^2} \Bigl(&  A_{ai} C_{k,ab} A_{bj} \omega_k
      + A_{ai} e^{\phi_a} A_{aj}  \psi_a \Bigr) -
    \\
\frac{1}{2\alpha^2}\Bigl( &
    e^{-\phi_c} C_{k;cb} A_{bi} C_{r;ce} A_{ej} \omega_r\omega_k
      + e^{\phi_c} A_ {ci} A_{cj}\psi_c\psi_c
     + A_{ci} C_{r;ce} A_{ej}\omega_r \psi_c
      +  C_{k;cb} A_{bi} A_{cj}\psi_c \omega_k \Bigr). \\
\end{split}
}
Rotation of ${\cal K}_{ij}$ gives
\labeq{T2}{
2\alpha^2 A_{ri} A_{sj} {\cal K}_{ij} 
=  \delta_{rs}e^{\phi_r} \Bigl( \psi \psi_r - \psi^2_r \Bigr) 
 - C_{n;rs} \omega_n \Bigl( \psi_r + \psi_s - \psi \Bigr)
 - e^{-\phi_c} C_{k;cr} C_{n;cs}\omega_n\omega_k. 
}
First term in \eqref{T2} has only diagonal, 
second only off-diagonal, and the third has both diagonal and off-diagonal 
non-zero elements.

Consider first a diagonal part of \eqref{d2gdt2-rot1}. Gathering diagonal 
contributions from the last four terms on the right-hand side of \eqref{d2gdt2-rot1} we obtain a set of evolution equations for deformation rates $\psi_i$,
\labeq{dpsidt}{
\begin{split}
&  
    \pd{\psi_i}{t} 
      - \pd{\ln \alpha}{t}\psi_i 
     = 2 \alpha^2 e^{-\phi_i} A_{in} A_{im} \left( 
              \frac{1}{\alpha}\nabla_n\nabla_m\alpha  
                  - R_{nm} \right) 
                         - \psi \psi_i
 + \Psi_{i;kn} \omega_n \omega_k +  e^{-\phi_i} A_{im} A_{in} {\cal S}_{mn}, \\
\end{split}
}
where 
\labeq{PSI}{
\Psi_{i;kn} 
= \left( e^{\phi_i - \phi_c} -  e^{\phi_c - \phi_i}\right) A_{im} A_{ir} B_{k;cm} B_{n;cr}, \quad \sum_i \Psi_{i;kn} = 0. 
}
%
%
%
Gathering off-diagonal
 contributions in 
 \eqref{d2gdt2-rot1}
we obtain 
\labeq{offdiag}{
\begin{split}
 C_{k;ij} & \left(  \pd{\omega_k}{t} 
                - \pd{\ln \alpha}{t} \omega_k 
                + \psi\omega_k 
          \right)
= \\
& 2\alpha A_{in} A_{jm} \nabla_n\nabla_m\alpha 
 -2 \alpha^2 A_{in} A_{jm}  R_{nm} + \Omega^1_{ij;nm}\omega_n\omega_m 
       + \Omega^2_{ij;nm} \psi_n\omega_m + A_{im} A_{jn} {\cal S}_{mn}, \\
\end{split}
}
where
\labeq{Omega1}{
\begin{split}
 \Omega^1_{ij;nm}  =  & 
    A_{it} A_{js} B_{n;ct} B_{m;cs} 
       ( e^{\phi_i + \phi_j -\phi_c} - e^{\phi_c} )  \\
 &         - B_{n;is} B_{m;js} ( e^{\phi_j}  + e^{\phi_i} ) 
  - e^{\phi_i} E_{mn;ik} A_{jk} 
 - e^{\phi_j} E_{mn;jk} A_{ik}  ,\\
\end{split}
} 
\labeq{Omega2}{
\Omega^2_{ij;nm} =  A_{ik} B_{m;jk} e^{\phi_j} ( \delta_{in} - \delta_{jn} )
                 + A_{jk} B_{m;ik} e^{\phi_i} ( \delta_{jn} - \delta_{in} ),
}
\labeq{}{
\Omega^1_{ij;nm} = \Omega^1_{ji;nm}, \quad \Omega^2_{ij;nm} = \Omega^2_{ji;nm}, \quad \Omega^2_{ii;nm} = 0.
}
Finally, solving equations \eqref{offdiag} we obtain
a set of equations for angular velocities $\omega_i$,
\labeq{domegadt}{
\begin{split}
& \pd{\omega_k}{t} - \pd{\ln \alpha}{t} \omega_k + \psi \omega_k = \\
& C^{-1}_{kr} \vert \epsilon_{rij} \vert 
         \Bigl( 2\alpha A_{in} A_{jm} \nabla_n\nabla_m\alpha
               -2 \alpha^2 A_{in} A_{jm} R_{nm} 
               + \Omega^1_{ij;nm}\omega_n\omega_m 
               + \Omega^2_{ij;nm} \psi_n\omega_m 
               + A_{im}A_{jn}{\cal S}_{mn}\Bigr). \\
\end{split}
}
Equations \eqref{dpsidt}, \eqref{domegadt} together with two equations
\eqref{psiomega} constitute an evolution part of an ADM system transformed 
to ESR variables.


Rewriting constraint equations is straightforward but the resulting formulas
are rather complicated and we do not present them here.
 To derive these formulas, one has to substitute expressions for 
the Riemann tensor \eqref{Ri},
 extrinsic curvature \eqref{Krot}, \eqref{Krot1}, \eqref{Krot2}, and 
Cristoffel symbols \eqref{Cr} into \eqref{ADMC}.
In rewriting the momentum constraints, one has also to differentiate
$K^i_j$ and $K$ with respect to $x^k$, which in turn requires 
differentiation of 
terms containing rotation matrices $A_{ij}$ and their first derivatives.
Expressions of second-order derivatives of $A_{ij}$ are given in \eqref{Ekr}.

For the energy constraint, one can also obtain a 
simple formula expressing $C_0$ through right-hand sides of the 
evolution equations for $\psi_i$. 
Summing up \eqref{dpsidt} and using \eqref{Krot2} we obtain
\labeq{dsumpsi}{
\begin{split}
\pd{\psi}{t} -  \pd{\ln\alpha}{t}\psi =    - & \alpha^2 \left(R + K^2 -
        \frac{1}{\alpha }  \gamma^{nm}   \nabla_n\nabla_m\alpha \right) ~+\\
                & 2\psi\gamma^{nm}\nabla_n\beta_m
                - \left(\gamma^{nm}\nabla_n\beta_m \right)^2
                + \half \gamma^{nm} {\cal S}_{nm}. \\
\end{split}
}
First and second terms in brackets in \eqref{dsumpsi} 
can be replaced using the
expression \eqref{ADMC} for the energy constraint, 
\labeq{dsumpsi1}{
\begin{split}
\pd{\psi}{t} -  \pd{\ln\alpha}{t}\psi =  - 
 & \alpha^2 \left(C_0 + K_{nm} K^{nm} -
        \frac{1}{\alpha}  \gamma^{nm}   \nabla_n\nabla_m\alpha \right) ~+ \\
            &     2\psi\gamma^{nm}\nabla_n\beta_m
                - \left(\gamma^{nm}\nabla_n\beta_m \right)^2
                + \half \gamma^{nm} {\cal S}_{nm}. \\
\end{split}
}
The latter equation can be inverted to obtain the energy constraint as
\labeq{C0}{
\begin{split} 
\alpha^2 C_0 = - & \frac{1}{2} \sum \pd{\psi_k}{t} + \pd{\ln\alpha}{t}\psi
  - \alpha^2 \left( K_{nm} K^{nm} -
        \frac{1}{\alpha}  \gamma^{nm} \nabla_n\nabla_m\alpha \right) ~+ \\
                & 2\psi\gamma^{nm}\nabla_n\beta_m
                - \left(\gamma^{nm}\nabla_n\beta_m \right)^2
                + \half \gamma^{nm} {\cal S}_{nm}. \\
\end{split}
}
We see that in ESR formulation the energy constraint depends 
linearly on time derivatives
of $\psi_i$. Setting $ C_0 = 0$ in \eqref{C0} will define 
a  plane in the three-dimensional space of $\pd{\psi_i}{t}$ in which 
the evolution of a constrained solution must occur, 
regardless of a choice of  gauge. The unit normal to the 
plane is a vector with components 
$\{\frac{1}{\sqrt{3}}, \frac{1}{\sqrt{3}}, \frac{1}{\sqrt{3}}\}$. 
A distance from the plane is given by the actual value of $-\alpha^2 C_0$.
Modifying \eqref{dpsidt} by adding the term $-\alpha^2 C_0 / \sqrt{3}$
to each of the right-hand sides will project  numerical evolution
onto a local $C_0=0$ plane, and may help with improving the accuracy of 
the evolution along a (generally curved) surface of constrained solutions.

We also note that differentiation of \eqref{Krot2} gives
\labeq{}{
\pd{K}{t} = - \frac{1}{\alpha} \left( \pd{\psi}{t} 
                                      - \psi \pd{\ln\alpha}{t}
                               \right) 
    + \pd{}{t} \left( \frac{\gamma^{nm}\nabla_n\beta_m}{\alpha} \right),
}
so that we can rewrite \eqref{dsumpsi}, assuming $C_0=0$,
 in a more familiar form as
\labeq{dsumpsi2}{
\pd{K}{t} =  \alpha  K_{nm} K^{nm}  
-  \gamma^{nm} \nabla_n \nabla_m \alpha ~+ ~{\rm "shift-dependent~terms"}.
}

\subsection{Case of degenerate eigenvalues}

As was mentioned in the beginning of the paper, decomposition
 \eqref{gamma1} is always possible but it is not unique if
 $\gamma_{ij}$ has degenerate eigenvalues. In case of two 
degenerate eigenvalues,  rotation angle in the direction orthogonal to 
 the corresponding eigenvectors
is arbitrary. In case of three  degenerate
eigenvalues all three rotation angles are arbitrary. For example,
a flat three-dimensional hypersurface with Cartesian coordinate system on it
 is a case of triply-degenerate eigenvalues. In all of these cases,
the matrix $C_{ij}$ in \eqref{C} becomes
degenerate and cannot be inverted. Equation 
 \eqref{domegadt} then cannot be used to determine $\pd{\omega_i}{t}$.
Below we describe how $\pd{\omega_i}{t}$ should be determined when 
eigenvalues are degenerate.

Consider first the case of triple degeneracy, 
$\phi_1 = \phi_2 = \phi_3$. Using \eqref{prop1} we can rewrite \eqref{Ci} as
\labeq{}{C_{k;ij} = A_{in} B_{k;jn} ( e^{\phi_j} - e^{\phi_i} ).
}
We see  that in a triply-degenerate case all three $C_{k;ij} \equiv 0$,
and \eqref{offdiag} becomes a system of three algebraic
equations
\labeq{offdiag3}{
\begin{split}
& 2\alpha A_{in} A_{jm} \nabla_n\nabla_m\alpha 
 -2 \alpha^2 A_{in} A_{jm}  R_{nm} + \Omega^1_{ij;nm}\omega_n\omega_m 
       + \Omega^2_{ij;nm} \psi_n\omega_m + A_{im} A_{jn} {\cal S}_{mn} = 0 \\
\end{split}
}
with respect to $\omega_i$. We assume that conditions for lapse $\alpha$ and 
shift $\beta_i$ are given, and that $\phi_i$, $\theta_i$, $\psi_i$, and 
$\omega_i$ 
are known on a
three-dimensional hypersurface as a result of
previous evolution. Thus, first, second, and last terms 
in \eqref{offdiag3} are known, and  $\Omega^1_{ij;nm}$ and  
$\Omega^2_{ij;nm}$ which depend on $\phi_i$ and $\theta_i$ 
are known as well.
Then \eqref{offdiag3} is a set of three
quadratic equations for $\omega_i$ which are automatically 
satisfied at a degenerate point by continuity of $\omega_i$. 
 Differentiation of
\eqref{offdiag3} with respect to $t$  will give us expressions for
$\pd{\omega_i}{t}$ which will depend on $\phi_i$, $\theta_i$, $\psi_i$, 
$\omega_i$ and on $\pd{\psi_i}{t}$. The latter values are given by
\eqref{dpsidt}. Thus we can uniquely
 determine $\pd{\omega_i}{t}$ as functions
of $\phi_i$, $\theta_i$, $\psi_i$, 
$\omega_i$, and their spatial derivatives, 
which was our original purpose. 

The case of two degenerate eigenvalues can be treated similarly.
Without loss of generality, assume that $\phi_1=\phi_2$. Then
all elements of symmetric matrices $C_{k;ij}$ are zero excluding
 $C_{k;13} = C_{k;31} \neq 0$ 
and $C_{k;23} = C_{k;32} \neq 0$. Equations \eqref{offdiag} then 
reduce to one
algebraic equation with respect to $\omega_i$,
\labeq{ADMNEW4-2}{
\begin{split}
& 2\alpha A_{1n} A_{2m} \nabla_n\nabla_m\alpha
               -2 \alpha^2 A_{1n} A_{2m} R_{nm} ~+ \\
               &  \Omega^1_{12;nm}\omega_n\omega_m 
               + \Omega^2_{12;nm} \psi_n\omega_m 
                  + A_{1n} A_{2m} {\cal S}_{nm} = 0, \\
\end{split}
}
plus two partial differential equations
\labeq{ADMNEW4-3}{
\begin{split}
 C_{k;13} \left(  \pd{\omega_k}{t} 
                - \pd{\ln \alpha}{t} \omega_k 
                + \psi\omega_k 
          \right)
= 
& 2\alpha A_{1n} A_{3m} \nabla_n\nabla_m\alpha 
 -2 \alpha^2 A_{1n} A_{3m}  R_{nm} ~+ \\
 & \Omega^1_{13;nm}\omega_n\omega_m +
   \Omega^2_{13;nm} \psi_n\omega_m + A_{1n} A_{3m} {\cal S}_{nm}, \\
\end{split}
}
\labeq{ADMNEW4-4}{
\begin{split}
 C_{k;23} & \left(  \pd{\omega_k}{t} 
                - \pd{\ln \alpha}{t} \omega_k 
                + \psi\omega_k 
          \right)
= 
 2\alpha A_{2n} A_{3m} \nabla_n\nabla_m\alpha 
 -2 \alpha^2 A_{2n} A_{3m}  R_{nm} ~+ \\ &
 \Omega^1_{23;nm}\omega_n\omega_m 
       + \Omega^2_{23;nm} \psi_n\omega_m + A_{2n} A_{3m} {\cal S}_{nm}. \\
\end{split}
}
Differentiation of \eqref{ADMNEW4-2} allows us to determine one time derivative
say, $\pd{\omega_3}{t}$ as a function of other variables and two other 
derivatives,  
$\pd{\omega_1}{t}$, $\pd{\omega_2}{t}$. 
Substitution of  $\pd{\omega_3}{t}$
into \eqref{ADMNEW4-3} and \eqref{ADMNEW4-4} will then give us two 
equations to explicitly determine $\pd{\omega_1}{t}$, $\pd{\omega_2}{t}$.
Using them we can find $\pd{\omega_3}{t}$.

A similar approach should be used in case of degenerate eigenvalues 
to determine SRE variables and their derivatives
from $\gamma_{ij}$ and its derivatives on the initial hypersurface. 
We see from the above consideration that starting
from degenerate initial conditions with arbitrary
rotation angles $\theta_i$ does not mean that rotation velocities $\omega_i$
and accelerations $\pd{\omega_i}{t}$ can also be set arbitrary.
 Angular velocities  must be determined from \eqref{offdiag3}, and they 
will depend
among other things on lapse and shift conditions that 
we wish to use for the evolution.

\section{Conclusions}

In this paper we formulated a tensorial 
exponential transformation of a three-dimensional metric of space-like
hypersurfaces embedded in a four-dimensional space time in terms of 
exponential stretch-rotation, or ESR  variables (Section 2).
We derived formulas relating derivatives of the metric 
with the corresponding derivatives of these variables.
A 3+1 system of Einstein's equations, formulated in terms of ESR variables, 
describes time evolution of the metric
at every point of a hypersurface 
 as a continuous stretch and rotation of a local cartesian 
coordinate system in a tangential space (Section 3). 

We want to mention that certain exponential transformation of 
variables were used
in GR before for special symmetric cases.
Such transformations sometime simplify
analytical operations, and final results may sometime be formulated in a more 
compact form \cite{EXP-BEFORE}. The transformation \eqref{gamma1}, \eqref{gamma1a}, \eqref{gamma2} 
considered in this paper
is the most general exponential transformation which 
can be carried out for a three-dimensional metric without assuming any 
symmetries. 

A potential advantage of the ESR formulation is
 a change it introduces
to the structure of right-hand sides of Einstein's equations by eliminating
contravariant four-dimensional 
 metric $g^{ab}$ as a multiplicative term in these equations.
For a scalar non-linear hyperbolic equation considered in \cite{KHOKHLOV-03},
the removal of a term $g^{-1}$, where $g$ is a scalar equivalent
 of the metric, lead to a dramatic improvement in long-term stability of
numerical integration. 
Consider right-hand sides of \eqref{dpsidt} and \eqref{domegadt}. Ricci tensor
$R_{ij}$ and other terms written using ESR variables consist of parts that
 either do 
not have 
multipliers $\propto e^{-\phi_i}$ or have multipliers of the type
$\propto e^{\phi_i-\phi_j}$. That is, they either do not have $\gamma^{ij}$
multipliers or have multiplier that are ratios of different eigenvalues of
$\gamma_{ij}$. In front of the right-hand sides There are
multipliers of the type $\alpha^2 e^{-\phi_i}$ 
in front of the right-hand sides. We must recall that the
lapse is a $g^{00}$ component of the four-dimensional metric $g_{ab}$ and 
thus having this multiplier is equivalent of having ratios of the
 zero-th (negative) eigenvalue to other three (positive) 
eigenvalues of a four-dimensional metric. We see that 
 the net result of transformation is 
replacement of $g^{ab}$  with ratios of four-metric eigenvalues. 
We  hope that the right-hand sides of the evolution equations
\eqref{dpsidt} and \eqref{domegadt} may behave better than the right-hand 
sides of the original ADM equations. 

We want to discuss now  differences between the ESR formulation of 
Einstein's equations and other versions of 3+1 formulations of GR.
Zelmanov \cite{ZELMANOV},
 in his version of a 3+1 decomposition, chooses 
a congruence of time-like lines (motion of local reference frames), 
and then at 
each event on a time-line draws a small element of a three-dimensional space
orthogonal to that line (a local three-dimensional reference frame). 
Einstein's equations in his formulation describe
physical accelerations, deformations, and rotation of these reference frames
(and physics of matter if space is not empty). There are no global 
three-dimensional slices embedded in a four-dimensional space-time in his
 formulation. In a classical ADM 3+1 formulation \cite{ADM} of GR and 
subsequent 3+1 formulations, a four-dimensional
space-time is split into a family of three-dimensional space-like slices and 
a universal "time" is introduced to label these slices. 
A global coordinate system exists on each slice, and a three-dimensional 
metric of slices evolves with time according to Einstein's equations and
pre-determined gauge conditions. In tetrad formulations of GR \cite{TETRAD},
a family of orthogonal 4-vectors is introduced and Einstein's equations are
formulated in terms of variables projected onto tetrad components.

The ESR formulation presented here is 
 closest to a standard 3+1 ADM formulation. 
However, instead of metric and 
its derivatives, it
uses variables which can be thought of in the following way.
A three-dimensional metric 
defines
an orthogonal {\it triad}
 of space-like vectors directed along the main axis of
 the metric tensor  at each point of a hypersurface. 
ESR variables $\phi_i$ and $\theta_i$ tell how a 
three-dimensional coordinate system on a hypersurface 
must be rotated and re-scaled at each point 
so that it 
 will locally coincide with a three-dimensional 
 Cartesian coordinate
system defined by the orthogonal triad at this point. The evolution of ESR 
variables is equivalent to the evolution of the three-dimensional metric and is
governed by Einstein's equations written in terms of these variables plus
the pre-defined gauge conditions. There is no orthogonal tetrad in this 
formulation.

From the above discussion it is clear that ESR 3+1 formulation presented here
can be modified and extended by introducing new variables 
such as spatial derivatives
of the metric, and by addition of various combination of constraints, similar 
to modifications introduced to  a standard ADM formulation before \cite{3+1}.
Hyperbolicity and stability properties of ESR system and its modifications
will be studied in subsequent publications.
We plan to use this new system for numerical solutions
of certain problems of GR.

\bigskip
\noindent
This work was supported in part by the NASA grant SPA-00-067, 
Danish Natural Science Research Council through grant No 94016535, 
Danmarks
Grundforskningsfond through its support for establishment of the 
Theoretical Astrophysics Center,
 and by the Naval Research Laboratory
through the Office of Naval Research. We thank Andrey Doroshkevich (TAC) 
and Kip
Thorne (Caltech) for useful discussions. 
I.D. thanks the Naval Research Laboratory,
A.K. thanks the
 Theoretical Astrophysics Center,  and both authors thank 
Caltech for hospitality
during their visits.


\end{document}